\RequirePackage{fix-cm}
\documentclass[natbib,smallextended]{svjour3}       
\smartqed  
\usepackage{graphicx}
\usepackage{amssymb}
 \journalname{International Journal of Theoretical Physics}
\begin{document}

\title{Behavior of Friedmann-Lema\^itre-Robertson-Walker Singularities }


\author{L. Fern\'andez-Jambrina
}


\institute{Matem\'atica Aplicada\\
Universidad Polit\'ecnica de Madrid\\
Arco de la Victoria 4\\
E-28040-Madrid, Spain
              \email{leonardo.fernandez@upm.es}       
}


\maketitle

\begin{abstract}
In \cite{stoica} a regularization procedure is suggested for
regularizing Big Bang singularities in
Friedmann-Lema\^itre-Robertson-Walker (FLRW) spacetimes. We argue 
that this procedure is only appliable to one case of Big Bang 
singularities and does not affect other types of singularities. 
\end{abstract}

\section*{}
\label{intro}
Singularities are present in most physically relevant cosmological
models \cite{} and challenge the validity of General Relativity as 
the theory of gravitation. It is conjectured that such singularities 
will be removed or appeased by a quantum theory of gravity or by 
corrections to Einstein's theory.

In General Relativity the universe is a spacetime endowed with a
metric and at cosmological scale it is assumed to be homogeneous and
isotropic. Hence the metric in coordinates $t,r,\theta,\phi$ is given 
by
\begin{equation}ds^2=-dt^2+a^2(t)\left(dr^2+ 
r^2\left(d\theta^2+\sin^2\theta\,
d\phi^2\right)\right),\label{metric}\end{equation}
with the usual ranges for spherical coordinates. The only free 
function is the scale factor $a(t)$.

We consider here just flat cosmological models since they appear to be 
favored by observations.

Assuming that the universe is filled by a perfect fluid with energy 
density $\rho$ and pressure $p$, the equations which relate them to 
the scale factor $a(t)$ are Friedmann's equations
\begin{equation}\label{flrw}\rho=
\frac{3\dot a^2}{a^2},\qquad 
p=-\frac{2\ddot a}{a}-\frac{\dot a^2}{a^2}.\end{equation}

The problem is solved if we know the equation of state for the fluid, 
$p=p(\rho)$ or equivalently the barotropic index $w=p/\rho$.

For instance, if $w$ is constant, we can integrate Friedmann's 
equations
\begin{equation}\label{scale}a(t)=a_{0}t^{2/3(1+w)},\end{equation}up to a constant $a_{0}$, obtaining 
power-law models. They are valid for epochs with nearly constant 
equation of state. For these models, the energy density and the 
pressure take the form
\begin{equation}\rho(t)=\frac{4}{3(1+w)^2t^2},\quad 
p(t)=\frac{4w}{3(1+w)^2t^2}.\end{equation}

We see that for these models both energy density and pressure blow up 
as $t^{-2}$ at $t=0$ in coordinate time. It is the Big Bang 
singularity.

For instance, in a radiation dominated epoch, $w=1/3$, the scale 
factor behaves as $t^{1/2}$, whereas in a pressureless dust dominated 
epoch, $w=0$, it behaves as $t^{2/3}$. The case $w=-1$ corresponds 
to a dark energy dominated epoch and has a different solution,
\[a(t)=a_{0}e^{\sqrt{\Lambda/3}t},\quad \rho=\Lambda=-p,\]
in terms of the cosmological constant $\Lambda$.

\cite{stoica} suggests that Big Bang singularities may be removed by 
replacing the energy density and the pressure by their 
\emph{densitized} versions,
\begin{equation}\label{regular} \tilde \rho=\rho\sqrt{-g},\qquad \tilde 
p=p\sqrt{-g},\end{equation} where $g=a(t)^{3}r^2\sin\theta$ is the determinant of 
the metric (\ref{metric}).

The key feature of the result is that the products $\rho a^3$, $pa^3$ 
are expected to be smooth instead of blowing up at Big Bang. 
According to \cite{stoica}:
\begin{theorem}If a is a smooth function, then the densities 
$\tilde\rho$, $\tilde p$ are smooth (and 
therefore nonsingular), even at moments $t_{0}$ when $a(t_{0}) = 0$.\end{theorem}

Confronting this result with cosmological power-law models, we notice 
that for them the products $\rho a^3$, $pa^3$ behave as $t^{-2w/(1+w)}$. 
Hence, $\tilde \rho$ and $\tilde p$ do not blow up at $t=0$ if and 
only if the exponent $-2w/(1+w)$ is positive, that is, if and only if 
$w\in [-1,0]$, that is, quintessence cosmological models. 

Classical equations of state fall out of this result: For them energy
conditions (\cite{HE}) impose that $w\in[0,1]$.  The only case for which
$\tilde\rho$, $\tilde p$ are regular at $t=0$ is the one of
presureless dust, $w=0$.  This is to be expected, since for these 
models the scale factor is not even a $C^1$ function, since behaves as a power 
of time $t$ with exponent lower than one.

Although they are not considered in \cite{stoica}, the same happens with 
Big Rip singularities (\cite{rip}). These are strong 
singularities (\cite{puiseux}) and correspond to power-law models with 
negative exponent, that is, $w<-1$.

There are also singularities with finite scale factor $a(0)$ 
(\cite{sudden}, \cite{suddenferlaz}). They are classified as Type II, 
III and IV in \cite{NOT} and V in \cite{wsing}. For them close to $t=0$ the scale factor 
behaves as $a(t)\simeq a_{0}+a_{1}t^\alpha$, with 
$\alpha>0$, and so the energy density and the pressure
\begin{equation}
\rho(t)\simeq 3a_{1}^2\alpha^2t^{2(\alpha-1)},\quad
p(t)\simeq 2a_{1}\alpha(1-\alpha)t^{\alpha-2},
\end{equation}
behave behave as their densitized versions. Multiplyling by $a(t)^3$ 
does no affect their regularity.

Completing the classification of cosmological singularities
(\cite{grand}), we find Grand Bang and Grand Rip singularities.  For
them the scale factor vanishes or blows up exponentially, $a(t)\simeq
a_{0}e^{a_{1}t^{-\alpha}}$, depending on the sign of $a_{1}$, negative
por Grand Bang and positive for Grand Rip.  In this case, both the
energy density and the pressure diverge as a power of coordinate time,
\[\rho(t)\simeq 3a_{1}^2\alpha^2t^{-2(\alpha+1)}, \quad
p(t)\simeq-3a_{1}^2\alpha^2t^{-2(\alpha+1)},\]
and hence multiplying by a vanishing exponential scale factor (Grand 
Bang) brings vanishing instead of diverging $\tilde\rho$ and $\tilde 
p$. For Grand Rip singularities (diverging exponential scale factor), 
$\tilde \rho$ and $\tilde p$ diverge.

Anyway, the use of non-diverging versions of energy density and 
pressure in Einstein's equations does not prevent the formation of 
singularities:

According to General Relativity, non-accelerated observers follow
causal geodesics on a spacetime endowed with a metric which is a
solution of Einstein's equations.  Geodesics can be parametrized using
as parameter the proper time $\tau$, which is defined by the relation
$d\tau^2=-ds^2$.

Singularities appear in the form of incomplete causal geodesics 
(\cite{HE}), that is, causal geodesics which cannnot be defined for 
all values of $\tau\in\mathbb{R}$.

In the case of FLRW spacetimes, there are two types of geodesics
(\cite{puiseux}): radial and comoving geodesics.

Parametrizations of radial geodesics are obtained as solutions of
 first order equations, \begin{equation} 
t'=\sqrt{-\varepsilon+\frac{P^2}{a^2}},\quad r'=\frac{P}{a^2},\qquad\theta'=0, \quad \phi'=0, \end{equation} where the prime
stands for derivation with respect to $\tau$.  The constant $P$ is a
conserved quantity of geodesic motion, which is the linear momentum of
the geodesic, and $\varepsilon$ takes the value one for timelike
geodesics and zero por lightlike geodesics.

If the conserved quantity $P$ is zero, we have timelike comoving geodesics. 
For these coordinate time is essentially proper time, since geodesics 
equations reduce to \begin{equation}t'=1,\qquad r'=0, \qquad \theta'=0, \qquad 
\phi'=0.\end{equation}

It is clear that regularizing the energy density and the pressure as 
in \cite{stoica} does not affect the formation of singularities 
according to the previous equations.

Even in the case of timelike comoving geodesics, we could think in 
principle that they are complete, since their parametrization is 
just $t=\tau+\mathrm{const}.$ for all values of the proper time 
$\tau$. But most curvature invariants, for instance the Ricci 
curvature  $R=-\mu+3p$, point out that the manifold does not include 
the locus $t=0$, since the curvature becomes infinite there, 
regardless of the regularization.

Summarizing, we have shown that the regularization of the energy
density and the pressure suggested by \cite{stoica} does not allow the
extension of the spacetime beyond the Big Bang singularity, since the
scale factor of the universe $a(t)$ is not affected by this change and
it is the only function determining the geometry of the universe
(\ref{metric}).  In fact, the scale factor (\ref{scale}) is not a
$C^1$ function at $t=0$ for Big Bang singularities and hence second order equations for it, such as
Friedmann's equations (\ref{flrw}), are ill-defined for $a(t)$.
Causal geodesics still reach the singularity at $t=0$ in finite proper
time, where they meet a curvature singularity, since curvature
invariants of the spacetime diverge as $t^{-2}$, pointing out that the
geometry of the universe is singular there, regardless of the
regularization.  Besides, the rescaling (\ref{regular}) does not
produce regular versions of the energy density and the pressure in the
case of Big Bang singularities, except for the case of a pressureless
dust equation of state.

%
%

\nocite{*}


\end{document}